\documentclass[a4paper]{jpconf}
\usepackage{graphicx}

\usepackage{iopams}
\usepackage{hyperref}

\hypersetup{ bookmarks=false, colorlinks=true, breaklinks=true, linkcolor=blue, menucolor=blue, urlcolor=blue}

\begin{document}
\title{CASCADE - a multi-layer Boron-10 neutron detection system}

\author{M K\"ohli$^1$, M Klein$^{1,5}$, F Allmendinger$^1$, A-K Perrevoort$^1$, T Schr{\"o}der$^{2}$, N Martin$^{2,3}$, C J Schmidt$^{4,5}$, and U Schmidt$^1$}
\ead{koehli@physi.uni-heidelberg.de}
\address{$^1$ Physikalisches Institut, Heidelberg University, Im Neuenheimer Feld 226, D-69120 Heidelberg, Germany}
\address{$^2$ Heinz Maier-Leibnitz Zentrum (MLZ) and Physik Department,  Technische Universit{\"a}t M{\"u}nchen, Lichtenbergstr. 1, 85748 Garching, Germany}
\address{$^3$ CEA, Centre de Saclay, DSM/IRAMIS/Laboratoire L{\'e}on Brillouin, 91191 Gif-sur-Yvette, France}
\address{$^4$ GSI Detector Laboratory, Planckstr. 1, 64291 Darmstadt, Germany}
\address{$^5$ CDT CASCADE Detector Technologies GmbH, Hans-Bunte-Str. 8$-$10, 69123 Heidelberg, Germany}

\begin{abstract}
The globally increased demand for helium-3 along with the limited availability of this gas calls for the development of alternative technologies for the large ESS instrumentation pool. We report on the CASCADE Project - a novel detection system, which has been developed for the purposes of neutron spin echo spectroscopy. It features 2D spatially resolved detection of thermal neutrons at high rates. The CASCADE detector is composed of a stack of solid $^{10}$B coated Gas Electron Multiplier foils, which serve both as a neutron converter and as an amplifier for the primary ionization deposited in the standard Argon-CO$_2$ counting gas environment. This multi-layer setup efficiently increases the detection efficiency and serves as a helium-3 alternative.
It has furthermore been possible to extract the signal of the charge traversing the stack to identify the very thin conversion layer of about 1 micrometer. This allows the precise determination of the time-of-flight, necessary for the application in MIEZE spin echo techniques. 
\end{abstract}

\section{Introduction}

The world of neutron detection has changed. Much of what once was established technology has been discarded. For them now substitutional ones have been presented. It began with production of tritium and peaked at the crisis of helium-3. Part of it was given to sciences for basic or applied research.
Part for the industry, explorating oil deep in the rocks. And the largest part was given to Homeland Security, which above all else demanded for it for the protection against hazards. After the stockpile was nearly exhausted in 2008~\cite{He3crisis}, concerns over the future supply~\cite{Fed09} critical to perspectives of the European Spallation Source~\cite{TdrESS} led to developments of replacement technologies, most of them adapted from particle physics.
The CASCADE neutron detection unit~\cite{Cascade} is such a new generation system, which was designed specifically for the purposes of Neutron Spin Echo (NSE) spectroscopy~\cite{Nse1972}. This method and its successors, Neutron Resonance Spin Echo (NRSE)~\cite{Nrse1987} and MIEZE (Modulation of IntEnsity by Zero Effort)~\cite{Nrse1992}, demanded a highly granular and time resolved detector to be operated efficiently at high rates. The CASCADE design is based on a combination of solid boron-10 coatings in several layers, modern gas amplification stages, a microstructured readout, multichannel ASICs (Application Specific Integrated Circuit) and FPGA (Field Programmable Gate Array) hardware triggered data acquisition. It is successfully in operation at the Forschungs-Neutronenquelle Heinz Maier-Leibnitz at the instruments RESEDA~\cite{ResedaJLSRF}~\cite{NrseReseda} and MIRA~\cite{MiraJLSRF}.
\begin{figure}[h]
\includegraphics[width=15pc]{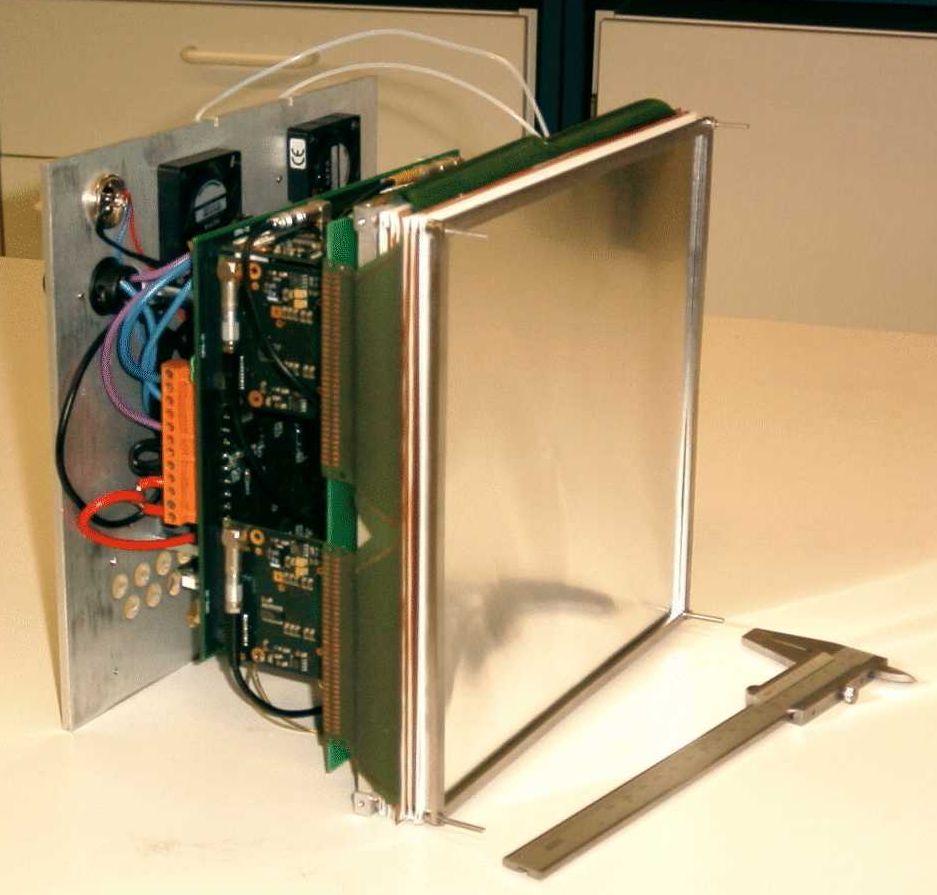}\hspace{2pc}%
\begin{minipage}[b]{20pc}\caption{\label{fig:cascadeNaked} The CASCADE detector (without housing and shielding): The 20x20\,cm$^2$ active detection volume behind the aluminum  window is composed of 6 layers of boron-10 coatings and the central double sided readout. It is flushed by a gas mixture of Argon:CO$_2$. The 5 CIPix charge sensitive preamplifiers with 64 channels each are mounted on a Virtex-2 data acquisition board. Additionally pulse height spectra can be acquired from each channel. The measurements can be transmitted optically as compressed raw data or stored onboard as histograms by the event analysis firmware.}
\end{minipage}
\end{figure}

\section{The CASCADE Detector}

The CASCADE Detector consists of several layers of boron-10 coated Gas Electron Multipliers (GEM)~\cite{SauliGem}. Thermal neutrons convert to charged ions via the reaction $^{10}\mathrm{B} + \mathrm{n} \rightarrow ^7\mathrm{Li} + ^4\mathrm{He}$ 
, which are emitted back to back at a total kinetic energy of up to 2.8\,MeV. These particles ionize the counting gas Argon:CO$_2$ (70:30-90:10). Subsequently the electrical charge is then projected towards a common readout. 


\begin{figure}[h]
\includegraphics[width=21pc]{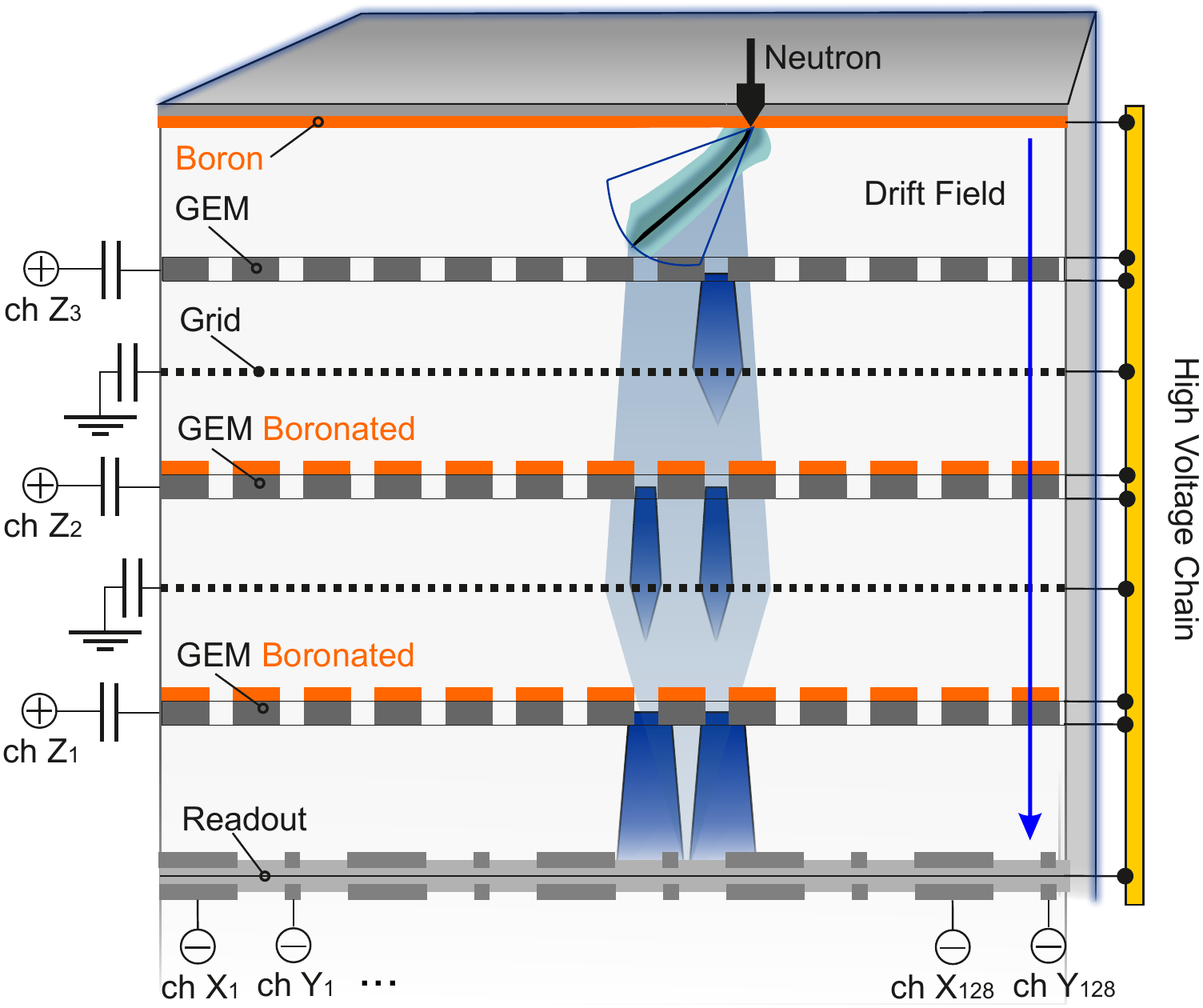}\hspace{2pc}%
\begin{minipage}[b]{14pc}\caption{\label{fig:schematicCascade} Schematic of the CASCADE detector of one half space: Thermal neutrons are converted in boron layers (orange). The conversion products ionize the counting gas and the electrons are projected towards a double sided readout of 128 channels in x and y each in crossed stripes of 1.57\,mm width. Additionally the signal induced by the charge traversing the stack allows the identification of the conversion layer. In order to decouple the GEMs electrically and to limit the generated charge, grids are inserted in-between each stage.  }
\end{minipage}
\end{figure}

The GEMs act as well as a substrate for the boron coating as they are 'charge transparent', which means that electron losses on the surface are compensated by gas amplification. GEMs are made of an insulator covered on both sides by copper. In etched holes of 0.05\,mm diameter a high field gradient between the conducting surfaces causes Townsend avalanches leading to a high gas gain. The opacity of 88\,\% therefore requires at least a real gain of approximately 10. Further features of GEMs, as compared to wire based systems, the good ion backflow suppression, the high rate capability and the decoupling of drift field and gas gain. The standard spacing of the holes of 0.18\,mm can be varied in the production process allowing for very high spatial resolution. In the CASCADE Detector presented here 6 layers divided into two half spaces are installed, see Fig \ref{fig:schematicCascade}, whereas in the top layer the drifts instead of the GEMs are coated. The coating thicknesses of 99\,\% enriched $^{10}$B for the RESEDA configuration according to the measurements of CDT CASCADE Detector Technologies GmbH are from top to bottom in micrometers: 1.5, 0.8, 1.0, 0.95, 0.8, 1.0 (originally 2.0 before 07/2013). The mean remaining energy of the conversion products to ionize the gas depends on the layer thickness. The charge is projected towards a common double sided readout, which is a printed circuit board of meander shaped crossed stripes.
Each side of 20\,cm is divided into 128 segments of 1.57\,mm width. Additionally to the 2D spatial readout it has been possible to measure the signal of the charge traversing the stack induced on the GEMs. For this reason there are metal meshes (grids) inserted in-between the layers. They decouple the GEMs electrically from each other and reduce the total amount of charge as the GEMs have to be operated at an effective gain larger than 1 to achieve clear signatures due to the large readout capacity and weak signals. The same ASIC is used for the GEM channels, which are called the z-coordinate. As the solid boron coatings are very thin, the conversion layer identification leads to a very high time-of-flight resolution.


A high integration density and charge sensitivity is achieved by using the 64-channel CIPix ASIC~\cite{cipixDipl}, which was developed on the basis of the HELIX128 chip as a readout amplifier for the inner tracker of the H1 detector at HERA.
It is operated at a constant frequency of 10\,MHz leading to a time resolution of 100\,ns. In total 5 CIPix boards are present in the system, whereas 4 are reserved for the (x,y) readout and one for the z coordinate. The digital signals are handled by a Virtex-2 FPGA board~\cite{Cascade}, which can either produce a zero-suppressed raw data output or it can analyze the data patterns by firmware algorithms in the triggered event mode. This generates and stores ready-to-use histograms, for example for stacking time-of-flight measurements. In the Spin Echo mode the Spin Echo group pattern can be followed synchronized to the coil frequency by means of a phase locked loop unit. For a MIEZE instrument this detector shows competitive performance to other NRSE setups. An efficiency of approximately 25\,\% for thermal neutrons and a spatial resolution of $\sigma=1.45$\,mm have been measured. At 16x sampling of the MIEZE modulated signal frequencies up to 625\,kHz can be used.

\section{Conversion products and their energy signatures}

\begin{figure}[h]
\includegraphics[width=16pc]{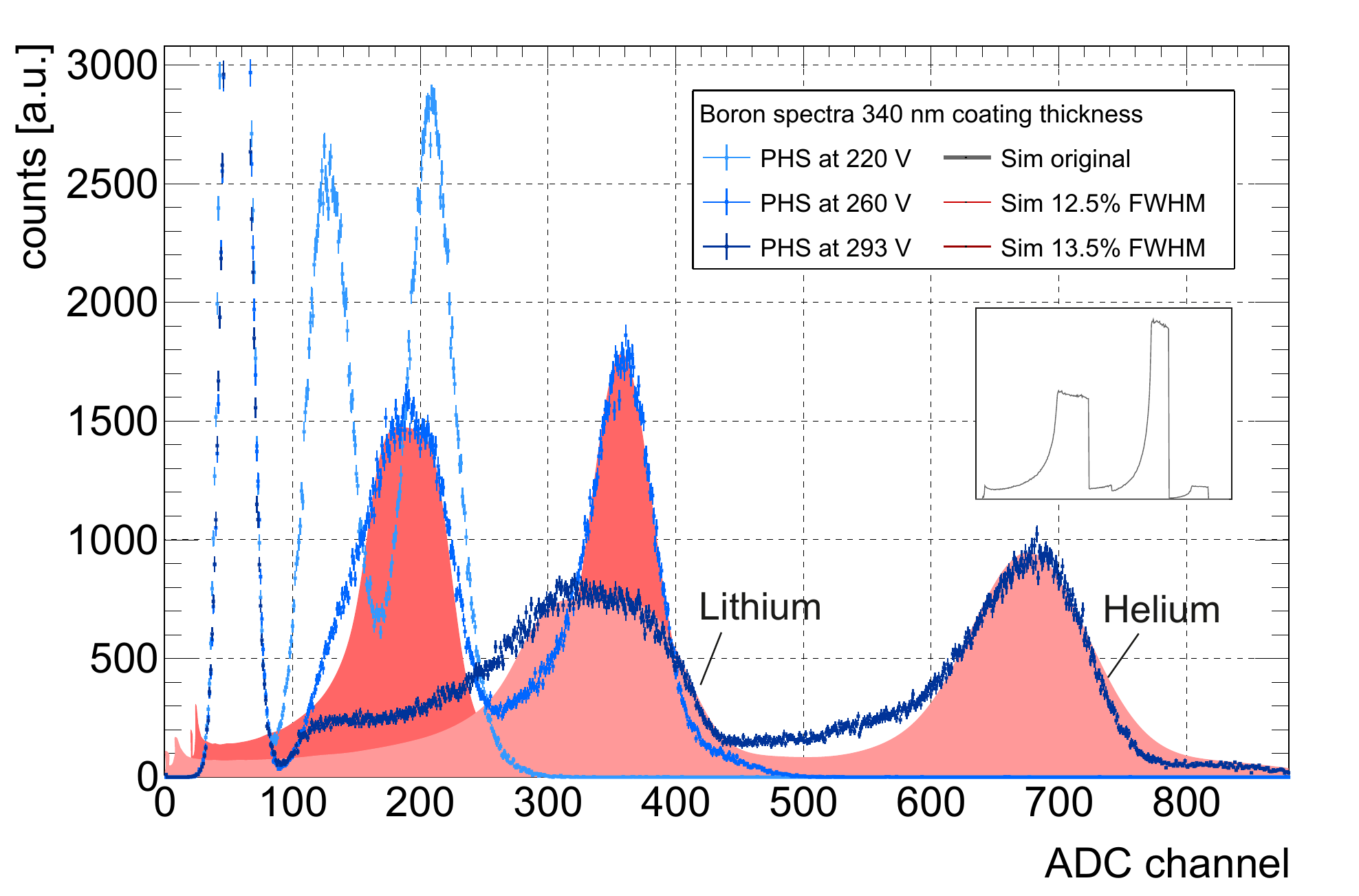}\hspace{2pc}%
\begin{minipage}[b]{19pc}\caption{\label{fig:PHSspectra} Pulse height spectra of the test detector, composed of a single coated GEM and the readout, at 3 different gas amplifications. The data is compared to results from the simulation of the system with the original spectrum (inlay) being convoluted by a Gaussian energy resolution function. }
\end{minipage}
\end{figure}

The thermal neutron capture leads the boron to fragment into a helium ion and the rest of the nucleus, which is referred to as a lithium ion. They either receive the full kinetic energy of 2.8\,MeV with a probability of 6.4\,\% or an additional photon of 0.48\,MeV is produced reducing the phase space accordingly. 
\begin{figure}[h]
\centering 
\includegraphics[width=0.99\linewidth]{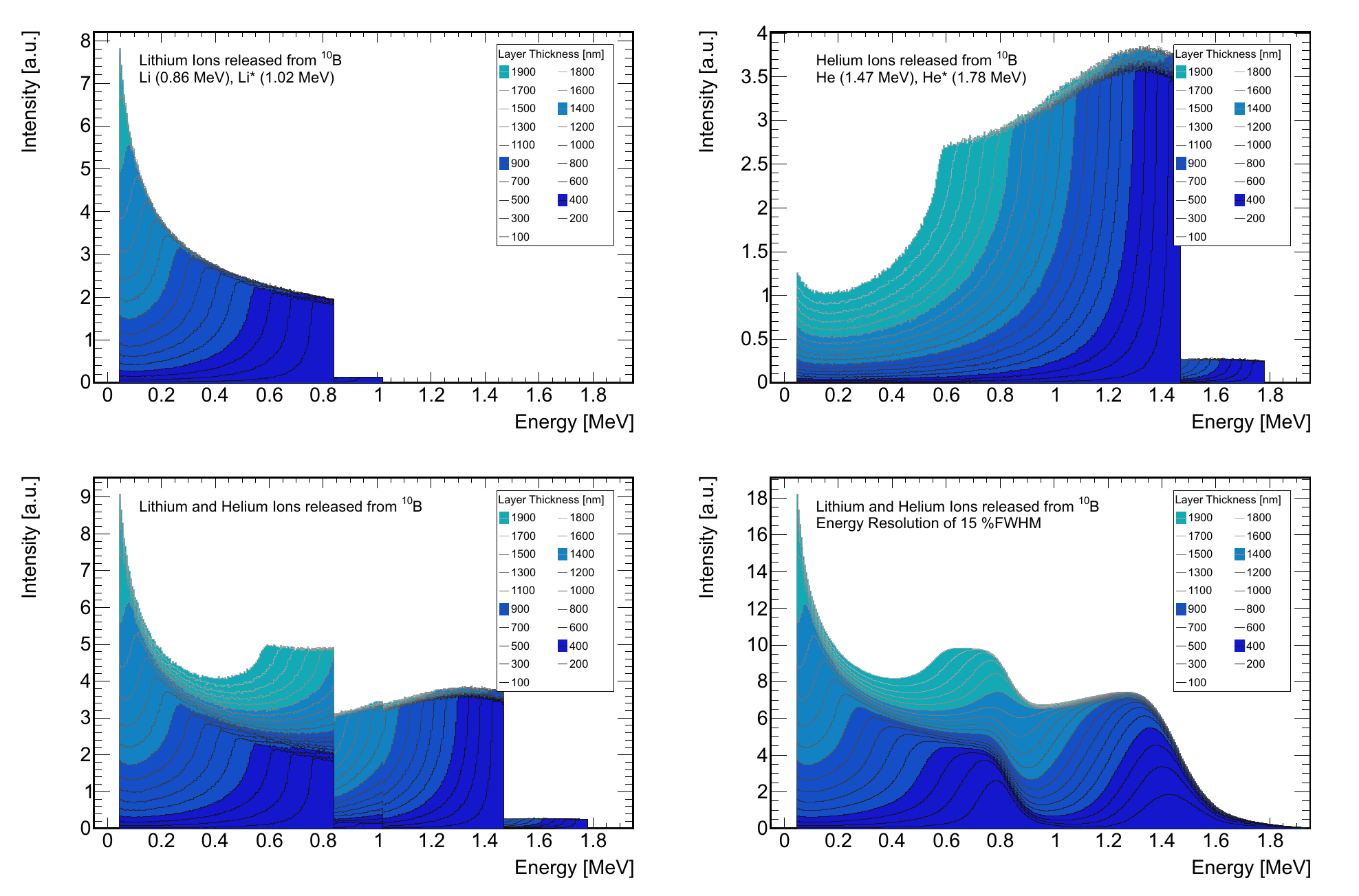}
\caption{Simulation of the energy spectra of conversion products leaving boron-10 layers of (100-1900)\,nm, colored in groups of 500\,nm. (top) Separated spectra by particle type: lithium (left) and helium (right). (bottom left) Combined spectra and (bottom right) combined spectra convoluted by an energy resolution of 15\,\% FWHM.}
\label{fig:boronSpectra}
\end{figure}
Whereas the helium ion can be detected easily, the lithium ion not only has less kinetic energy due to its higher mass but also the energy loss in the medium is higher. Therefore the range of these ions in boron is less than half compared to the 4\,$\mu$m of helium. Simulations of the energy spectra of the particles after leaving the converter were done by the Monte-Carlo code URANOS (presented in~\cite{myself}), which is linked to the stopping range calculations of SRIM
. The results are shown in Fig. \ref{fig:boronSpectra} for thicknesses of (100-1900)\,nm. 
Due to the stopping power the spectra extend towards lower energies for increasing coating layers. The helium ion signature can be detected far above 2000\,nm, whereas the lithium ion does not contribute any more to the efficiency beyond 1500\,nm. Additionally to the work presented here, analytical descriptions of the spectra have been presented in~\cite{Kle00} and calculations focusing on the efficiency of multi-layer systems have been proposed in~\cite{b10filmCalc}. For our specific case we measured these spectra in a test detector with a single boron coated GEM and a large drift volume. Fig. \ref{fig:PHSspectra} shows as an example the pulse height spectra for three different gas gains and in comparison the simulated spectra of Fig. \ref{fig:boronSpectra}, which have been convoluted by a Gaussian resolution function of the obtained peak full width half maximum. 

\begin{figure}[h]
\centering 
\includegraphics[width=\linewidth]{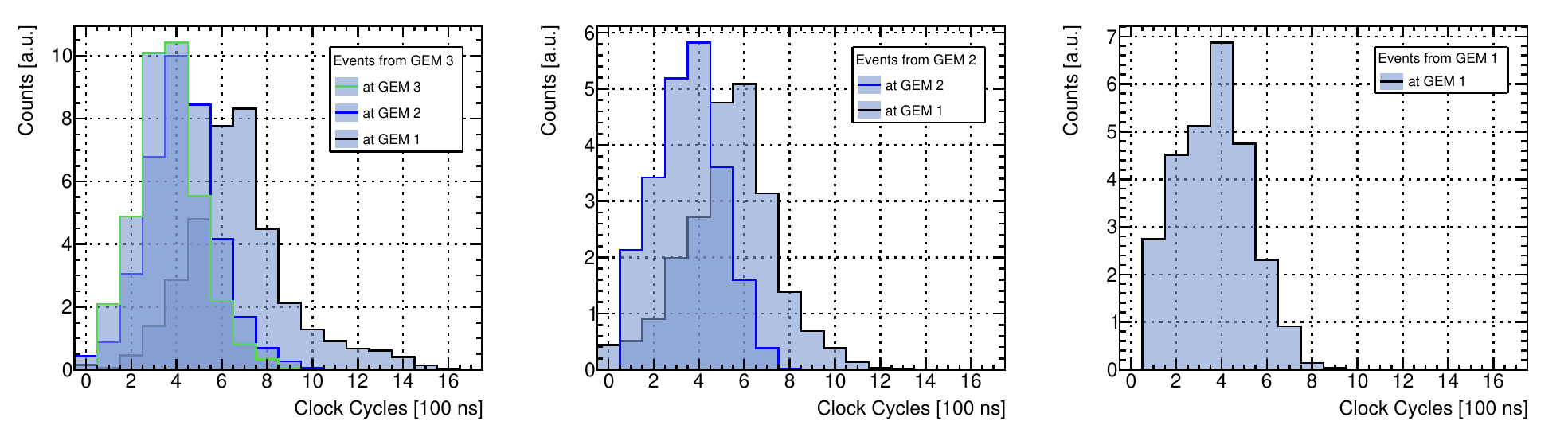}
\caption{Raw data spectra in units of clock cycles to read out the z-channels of the charge traversing the stack at a gain voltage of 350\,V per GEM of the RESEDA detector. GEM 3 denotes the layer next to the drift, the charge of a neutron event can traverse the whole stack and is measured in three layers. GEM 2 and GEM 1 are located in drift direction towards the readout. }\label{fig:rawDataSpectra}
\end{figure}

The gas ionization energy depositions here are slightly shifted downwards as the tracks emerging from the GEM are partly tallied by projection onto the surface instead of into a hole. In the CASCADE detector these spectra will be tallied approximately at 500\,keV of energy deposited in the gas. The track length of the conversion products of about 10\,mm in Argon:CO$_2$ is much higher than the available space between the layers of (2-4)\,mm. 
\begin{figure}[h]
\includegraphics[width=19pc]{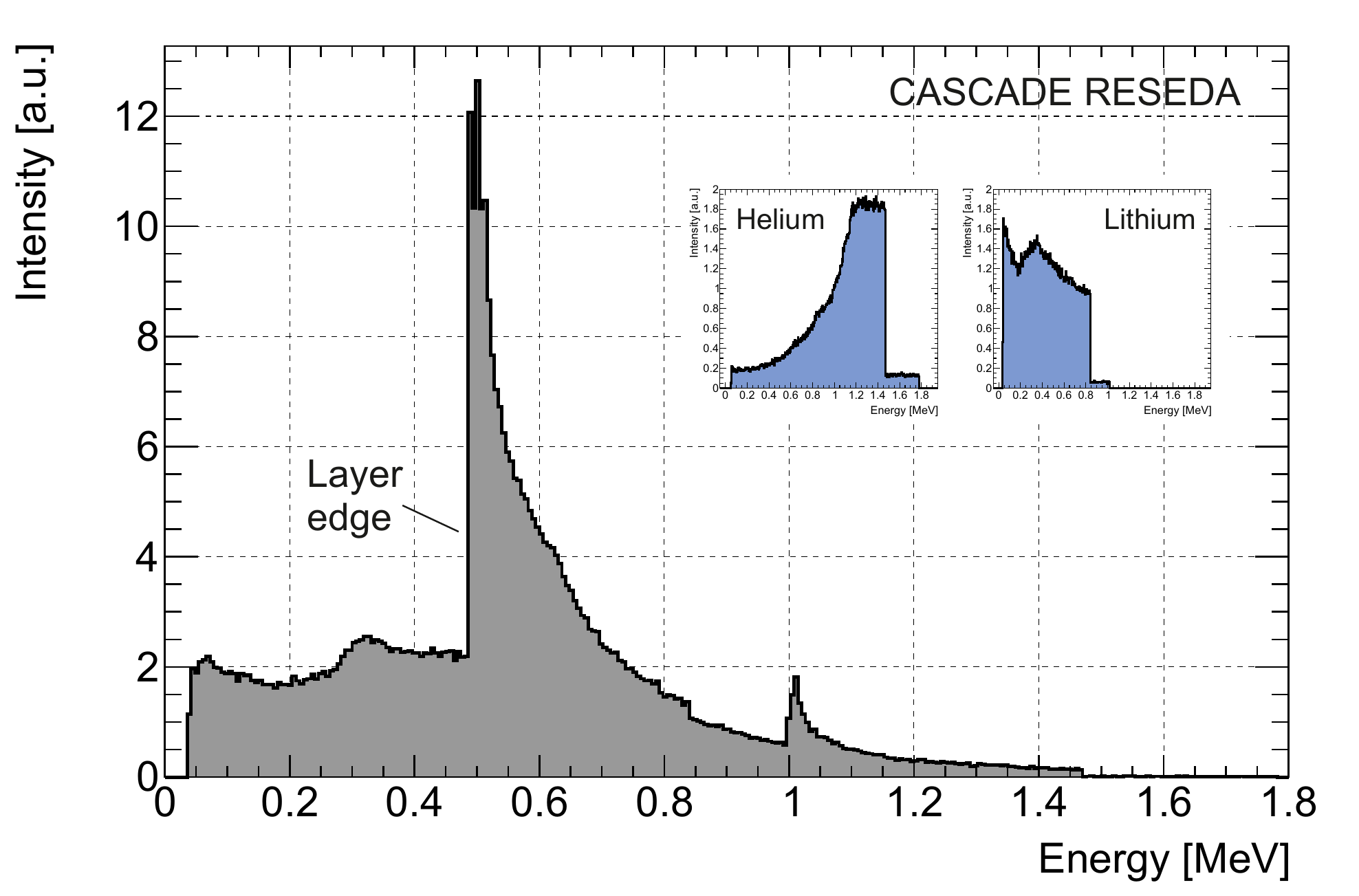}\hspace{2pc}%
\begin{minipage}[b]{16pc}\caption{\label{fig:PHSspectraCascade} Simulation of the gas energy deposition of the conversion products for a CASCADE detector in the RESEDA configuration. The inlays show the energy of the particles escaping the boron coatings. As different layer thicknesses are present in the detector, the distributions are not uniform. The geometry of each layer limits the track length and therefore the available energy.}
\end{minipage}
\end{figure}
In Fig. \ref{fig:PHSspectraCascade} a simulation of the cumulated energy deposited in the detector at RESEDA before 2013 is shown. The tail below 200\,keV comes from the thicker coatings on the front and back. For calculating the total neutrons measured additionally a detection threshold of (50-100)\,keV has to be considered. This especially has a non negligible effect as there is a pile up in the spectra on low energies for larger converter thicknesses. 
The digitized signals of the CIPix ASIC do not feature a simultaneous time over threshold (TOT) or pulse height measurement, instead the charge sensitive channels require a certain time to read out a specific amount of charge. For this reason it takes several clock cycles of 100\,ns to process signals present in the detector. This allows for a pseudo-TOT measurement of the charge, which is primarily important for the z-channels as the firmware algorithms can for example identify crosstalk between the layers based on this information. In normal operation the signals of the RESEDA detector as like depicted in Fig. \ref{fig:rawDataSpectra}. The panel shows the signal lengths of charge propagating through the stack for each of the three layers. 

The left graph shows neutron events originating from the boron coating of the drift cathode.  As the effective gas gain is larger than 1, the amount of charge increases each layer. It has to be noted that the average pseudo-TOT is (40-50)\,ns and some events take up to 150\,ns. A simple (x,y,z) correlation would limit the rate capability. For this reason the firmware algorithm analyzes the spatial and temporal structure of the data and dynamically masks active channels. As a result the long events in the time distribution can be separated and their effect on the rate capability marginalized. Furthermore as far as two consecutive neutrons do not convert in the same layer their origin can be effectively reconstructed well below the average pseudo-TOT.

\section{The detector in Spin Echo mode}

\begin{figure}[h]
\includegraphics[width=20pc]{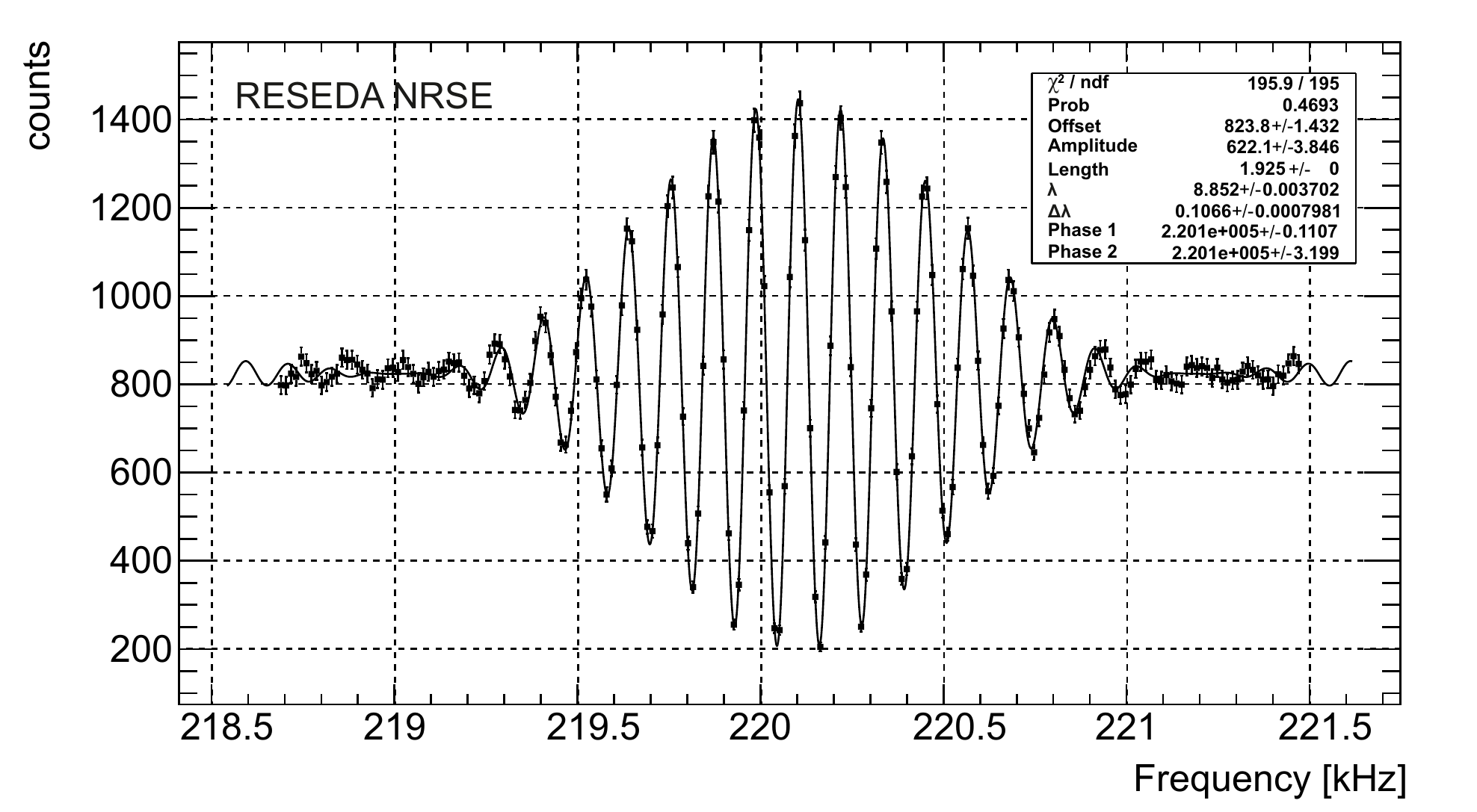}\hspace{2pc}%
\begin{minipage}[b]{15pc}\caption{\label{fig:nrseExample} Example of a Neutron Resonance Spin Echo (NRSE) Group measured by the CASCADE detector at RESEDA. By changing the frequency the interference pattern is moved spatially. At 8.85\,\AA\ and $\Delta\lambda/\lambda \approx 11$\,\% the polarization reaches 75\,\%. The color scale is normalized to the maximum intensity.}
\end{minipage}
\end{figure}

Spin Echo measurements, especially MIEZE techniques, require a very high spatial and time resolution. The oscillating interference pattern challenges detector technologies, as the Spin Echo group, which is depicted in the example in Fig. \ref{fig:nrseExample}, can have an extension in space in the order of only millimeters and has to be sampled in the time domain in the order of MHz. Recent user based measurements for example can be found in~\cite{Mieze17TReseda}.

\begin{figure}[h]
\centering 
\includegraphics[width=0.995\linewidth]{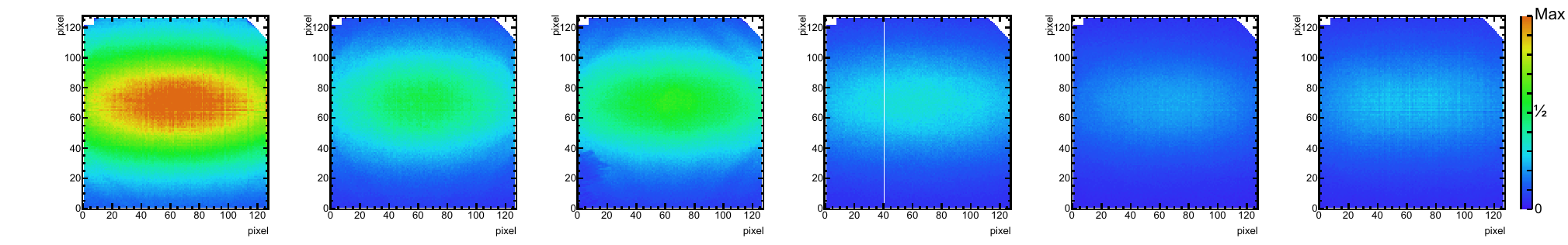}
\caption{Intensity distribution separated by GEM channel (top layer left, bottom layer right). This MIEZE measurement at RESEDA was conducted at a frequency of 53.5\,kHz and a wavelength of 8.05\,\AA .}
\label{fig:Nachtmessung}
\end{figure}

The multi-layer performance we want to illustrate by the capability of the detector to characterize itself in a neutron interference measurement. Whereas the data from Fig. \ref{fig:nrseExample} was taken in a standard NRSE setup, the following measurements were conducted in the MIEZE configuration. A thin graphite resolution sample was placed in the beam generating a homogenous illumination at the detector position. The intensity distribution integrated over the duration of the measurement is shown in Fig. \ref{fig:Nachtmessung} for each layer separately. The relative contribution of each channel depends on the boron thickness and the wavelength of the beam. As the cold part of the spectrum was used, most of the neutrons convert in the first layer.

\begin{figure}[h]
\centering 
\includegraphics[width=0.995\linewidth]{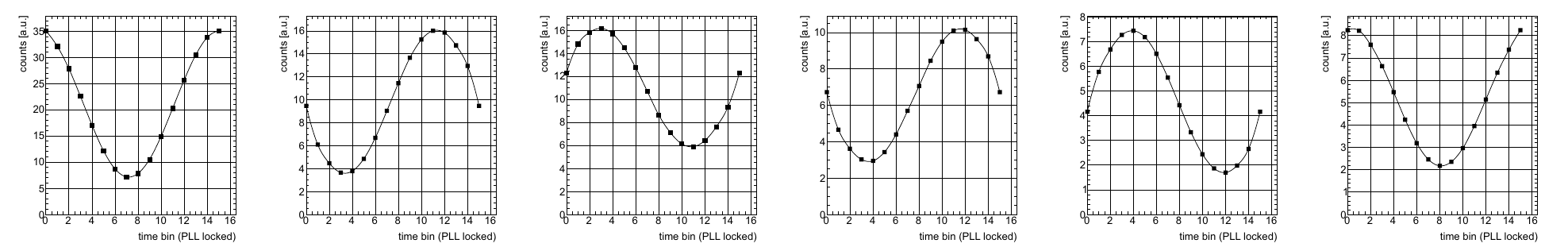}
\caption{Measurement of the intensity distribution for each GEM channel summed over a region of 9 pixels. The MIEZE period length of 18.7\,$\mu$s is oversampled by a PLL. The full cycle is divided into 16 subchannels. While the single point error is far below 1\,\%, the sine fit describe the data very well giving a reduced $\chi^2$ of approximately 1.}
\label{fig:NachtmessungSine}
\end{figure}

In a MIEZE measurement the intensity distribution oscillates in space and time as the neutron phase travels through the detector. In the setup presented the frequency of the polarization change of 53.5\,kHz is fed into the PLL and oversampled 16 times\footnote{To avoid jitter effects of the electronics, which would wash out the phase locked sine function and therefore reduce the measured polarization, the period is sampled in 16+1 time bins and afterwards recorrected. This relaxes the necessary precision to half a time bin on the interval boundaries and therefore reduces noise effects.}, which results in the detector following phase locked to the instrument at a readout frequency of 856\,kHz. In every pixel on every layer of Fig. \ref{fig:Nachtmessung} the oscillating interference pattern of 9.2\,mm wavelength can be measured. As an example Fig. \ref{fig:NachtmessungSine} zooms into one small region and shows the time channel of each layer.

\begin{figure}[h]
\centering 
\includegraphics[width=0.995\linewidth]{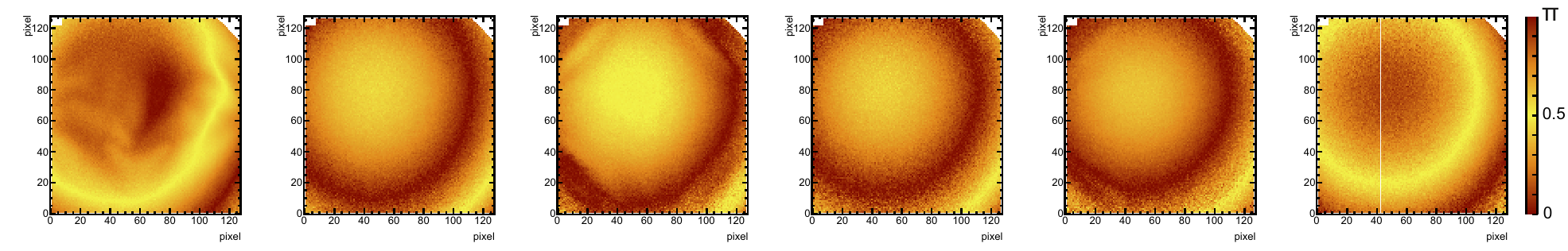}
\caption{Phases of the neutron beam separated by GEM channel (top layer left, bottom layer right). This MIEZE measurement at RESEDA was conducted at a frequency of 53.5\,kHz and a wavelength of 8.05\,\AA . The color code scales half a period from 0 to $\pi$, which equals 9.34\,$\mu$s or 4.6\,mm. The inner layers distances are approximately half a period.  }
\label{fig:Nachtmessung2}
\end{figure}

The resulting temporal sine function of the interference pattern is measured precisely. The following panel Fig. \ref{fig:Nachtmessung2} shows the phase distribution across the detector at one specific point of time. The oscillation period can be followed through the layers in every pixel. This evidently shows that high spatial and time resolution is necessary to conduct this type of experiments.

\begin{figure}[h]
\includegraphics[width=13pc]{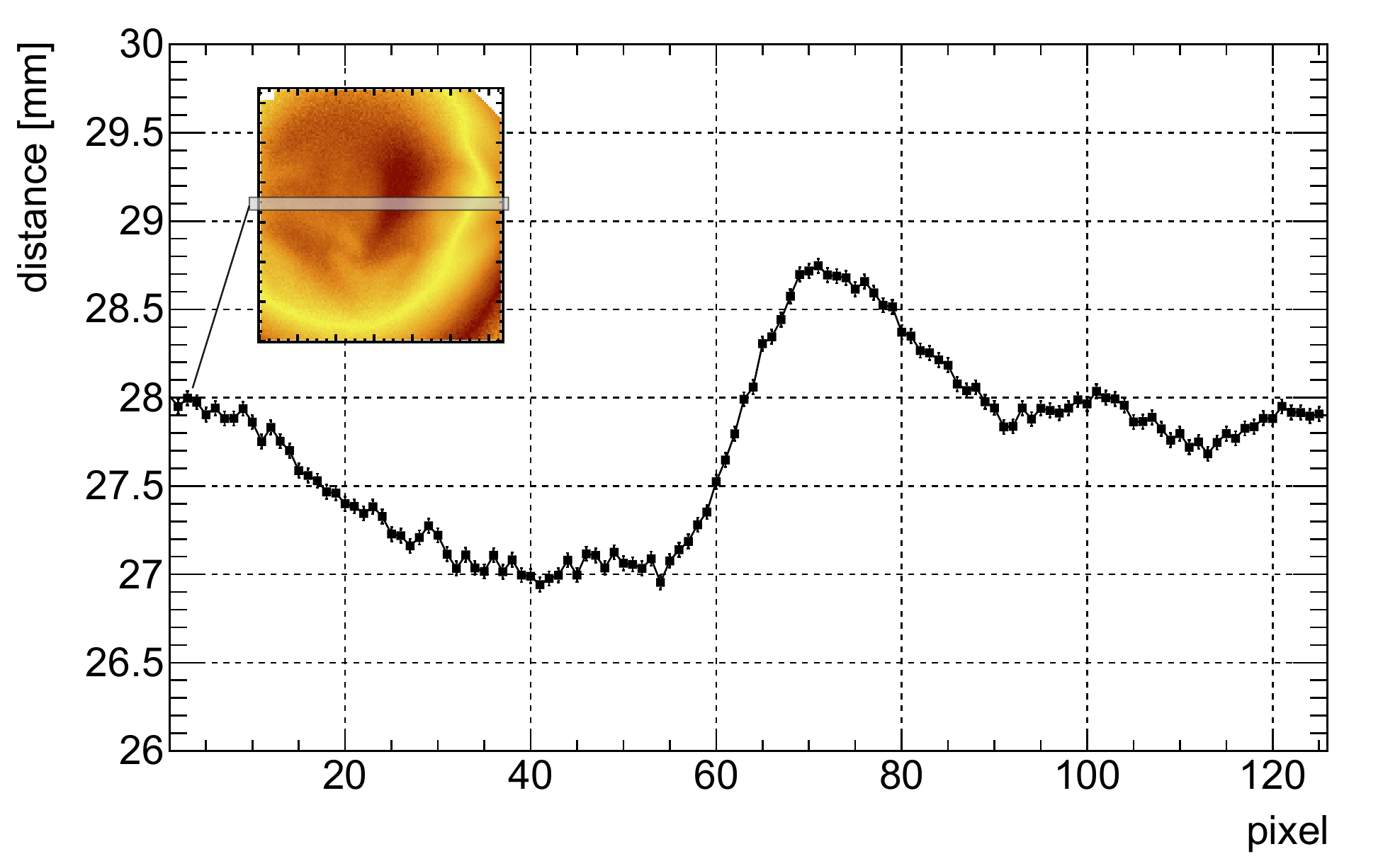}\hspace{2pc}%
\begin{minipage}[b]{22pc}\caption{\label{fig:profile} Profile view of the top to bottom displacement measurement by the neutron phase front. The cut view through the middle part of the detector depicts the bump on the drift cathode. The total displacement of 1.7\,mm can be determined down to a precision of 0.1\,mm.}
\end{minipage}
\end{figure}

We as well demonstrate how by an incidentally received bump on the front the detector can be aligned in a Spin Echo type of experiment. As the neutron speed and oscillation period are well known by the instrument setup, the phase is used to characterize the layers of the detection system spatially in respect to the beam axis and towards each other. 
As an example the front to back distance of the first and the last conversion layer is depicted in Fig. \ref{fig:profile}. As the bottom drift cathode is straight and therefore acts as a reference the displacement is attributed to the top layer. We can determine the bump depth of 1.7\,mm with a z-axis resolution of 0.1\,mm. For a shorter oscillation period length than what is available for this accidental measurement the analysis can easily be improved by a factor of 10. Using this time of flight principle it is possible to physically characterize and align the detector itself by the precision of the neutron interference pattern. 

\section{Conclusion}

The CASCADE detector has become an established device for Spin Echo measurements. Besides the high precision requirements of a NRSE instrument the detector proposes a design for a helium-3 replacement system by technology transfer from developments of particle physics. This paper presents the working multi-layer CASCADE implementation at RESEDA in the technical realization, electronics and firmware. The use of Monte-Carlo techniques to model the detector is necessary to understand the details of the underlying physics. In exemplary measurements we show the multi-layer operation in the Spin Echo mode by tracking the neutron interference pattern spatially and time resolved. The detector can be operated up to a rate capability of 2\,MHz on its active surface of 20x20\,cm$^2$. It features a time resolution of 100\,ns and the spatial resolution has been determined to $\sigma = 1.45$\,mm. The absolute detection efficiency of the detector at RESEDA is 21\,\% for thermal neutrons or 47\,\% at 5.5\,\AA. More details on the characteristics will be presented in further publications.

\section{Acknowledgments}

MK thanks W. H\"au\ss ler for support during his active period as responsible at the RESEDA instrument at the FRM II and T. Weber as well as R. Georgii. 
This work has been funded by the German Federal Ministry of Education and Research (BMBF) for the project ''Neutron Detectors for the MIEZE method'', grant identifier 05K10VHA.

\section*{References}

\bibliography{Literature}

\end{document}